\documentclass[aps,prl,twocolumn,superscriptaddress,floatfix]{revtex4}
\usepackage{graphicx}% Include figure files
\usepackage{color}

\begin{document}

%\preprint{in preparation}

\title{Quenching of stimulated Raman scattering in the kinetic regime by external magnetic fields}
% Force line breaks with \\

\author{B.~J.~Winjum}
\affiliation{Institute for Digital Research and Education, University of California Los Angeles, Los Angeles, CA 90095}
\affiliation{Particle-in-Cell and Kinetic Simulation Software Center, University of California Los Angeles, Los Angeles, CA 90095}
\author{F.~S.~Tsung}
\affiliation{Particle-in-Cell and Kinetic Simulation Software Center, University of California Los Angeles, Los Angeles, CA 90095}
\affiliation{Department of Physics \& Astronomy, University of California Los Angeles, Los Angeles, CA 90095}
\author{W.~B.~Mori}
\affiliation{Institute for Digital Research and Education, University of California Los Angeles, Los Angeles, CA 90095}
\affiliation{Particle-in-Cell and Kinetic Simulation Software Center, University of California Los Angeles, Los Angeles, CA 90095}
\affiliation{Department of Physics \& Astronomy, University of California Los Angeles, Los Angeles, CA 90095}
\affiliation{Department of Electrical Engineering, University of California Los Angeles, Los Angeles, CA 90095}

\date{February 16, 2018}% It is always 
%\today%, today,
             %  but any date may be explicitly specified

%\pacs{52.38.Bv, 52.35.Mw, 52.35.Fp, 52.65.-y}%PACS, the Physics and Astronomy
                             % Classification Scheme.
%\keywords{Suggested keywords}%Use showkeys class option if keyword
                              %display desired

\begin{abstract}
We show via particle-in-cell simulations that small normalized magnetic fields ($\omega_c/\omega_p \ll 1$) applied perpendicularly to a light wave can significantly modify the evolution of backward stimulated Raman scattering (SRS) in the kinetic regime. The presence of the magnetic field increases the threshold for kinetic inflation and decreases the amount of reflectivity when SRS is driven significantly above threshold. Analysis indicates this arises because trapped electrons are accelerated as they surf across the wave, leading to the continual dissipation of the electron plasma waves over a wider range of wave amplitudes. The simulation parameters are directly relevant for SRS in inertial confinement fusion devices and indicate that approximately 30 Tesla magnetic fields might significantly reduce SRS backscatter.
\end{abstract}

\maketitle

The nonlinear damping of electron plasma waves (EPWs) propagating in a magnetic field is a topic of fundamental interest. However, it has not received much attention in relation to topics such as inertial confinement fusion (ICF) \cite{betti:16} where nonlinear EPWs might be very influential but where magnetic fields (applied or self-generated) are not assumed to be important because they are relatively small.  In ICF, the driving laser light can decay via stimulated Raman scattering (SRS) into scattered light and EPWs.  SRS can reflect a significant amount of the driving laser energy and the resulting EPWs can generate non-thermal tails of energetic electrons that potentially preheat the fuel.  For kinetic SRS, electrons that interact resonantly with the EPW can reduce its damping rate (leading to kinetic inflation \cite{vu:07,ellis:12}), alter its frequency, bend its wavefronts, change its envelope shape, and couple it with other plasma modes (e.g., \cite{yin:07,yin:08,yin:13,winjum:10a,winjum:10b}).  

Several authors have recently investigated how an external magnetic field ($B_{ext}$) could potentially impact ICF performance through its effect on implosion dynamics, fusion reactivity, and hot electron propagation \cite{perkins:13, strozzi:15,strozzi:15jpp}, and several authors have pointed explicitly to using $B_{ext}$ as a method to indirectly decrease SRS.  Montgomery \textit{et al.} \cite{montgomery:15} have shown that 7.5T B fields can increase the electron temperature in hohlraums and hypothesized that $B_{ext}$ will thereby increase the Landau damping rate and decrease SRS growth.  Yin \textit{et al.} have argued that applying $B_{ext}$ aligned with the laser propagation direction (and the daughter EPW propagation direction) will limit the transverse motion of resonant electrons and reduce collective cascades of multi-speckle SRS \cite{yin:13}.  Barth \textit{et al.} have shown in simulations that Faraday rotation can disrupt the action of laser-plasma interactions \cite{barth:16}.  

Another motivation to use $B_{ext}$ that has not been investigated is its ability to alter the resonant wave-particle interactions and thereby directly increase the EPW damping.  In this Letter, we show for the first time that applying an external B-field perpendicularly to an incident laser beam can quench SRS activity.  This limiting effect on SRS is due to the damping of nonlinear electron plasma waves propagating across an external B-field.  The amplitude of magnetic fields required to significantly reduce SRS are dependent on the parameter regime, though for the cases considered here are on the order of 10's of Tesla and within the parameter ranges recently studied by other authors \cite{montgomery:15,perkins:13,strozzi:15,strozzi:15jpp}.

We consider parameters where $\bar \omega_c\equiv \omega_c/\omega_p \ll 1$, where $\omega_c$ and $\omega_p$ are the electron cyclotron and plasma  frequencies, respectively ($\bar \omega_c = 3.3 \times 10^6 B_{Tesla}/(n_{cm^{-3}})^{1/2}$).  In this regime, the real frequency of the plasma wave is essentially its unmagnetized value and Faraday rotation of the light waves is negligible. We also consider situations in which the bounce frequency $\omega_B \equiv \sqrt{eE_0k/m}$ is larger than the Landau damping rate $\gamma_{LD}$, where $E_0$ and $k$ are the EPW's amplitude and wavenumber and we note that $\omega_B/\omega_p \equiv \sqrt{\frac{eE_0k}{m\omega_p^2}} \cong \sqrt{\epsilon}$, where $\epsilon$ is $E_0$ normalized to the cold wavebreaking value \cite{dawson:59}.  Under these conditions, a plasma wave will evolve toward undamped modes after several bounce times  in an unmagnetized plasma. For such situations, Sagdeev and Shapiro \cite{sagdeev:73} and Dawson \textit{et al.} \cite{dawson:83} have shown that the initial damping of the wave, the evolution of the wave after several bounces, and its long time evolution after many bounce times are all profoundly effected by even weak fields, due to the fact that trapped electrons (those moving near the phase velocity $v_{\phi}$ of the wave) in an average sense all get accelerated perpendicularly across the wave front, continually extracting energy from it. For an EPW of the form $\vec E = E_0 \sin (kx-\omega t) \hat x$ propagating perpendicular to a field $\vec B=B_0\hat z $, the equations of motion for an electron in the wave frame are:  $\ddot{v}'_x + \omega_B'^2v_x' = -\omega_c^2v_{\phi}$ and $\dot{v}_y = \omega_c v_x$, where $\omega'_B=\sqrt{\omega_c^2+\omega_B ^2\cos(kx)}$ and primed quantities ($'$) are defined in the wave frame. Although not shown here, a variety of trajectories can be seen by solving these equations. The exact trajectory in real (and velocity) space depends on the initial phase of the electron. However, as shown in Dawson \textit{et al.}, in a large amplitude EPW a typical particle will execute bounce motion at roughly the modified frequency $\omega'_B$ while it is accelerated (deflected) transversely across the wave front. As this occurs, the resulting $v_y B_0$ force causes the electron to slowly drift backwards in the wave frame. In 1D, and if the electron starts near rest at the bottom of the wave's potential ($-e\phi$), i.e., at a zero of the electric field where its slope is positive, the particle will continue to execute this modified bounce motion until the $E_x + \frac{v_y}{c}B_0$ force vanishes, at which time the electron will be ejected with $v_y=c\frac{E_x}{B_0}$.  However, if the electron starts at different locations in the potential well, or if it begins with a large $v_x$ as might occur if the trapping width is large, then it can exit with a value of $v_y$ more than an order of magnitude less than $c\frac{E_x}{B_0}$. 

We stress, however, that for SRS in high energy density plasmas, there is a spectrum of plasma waves, the plasma wave amplitudes and phases are continually changing, and relativistic effects can be important.  For ICF parameters, $\frac{E_x}{B_0} \gg 1$, but relativistic corrections and additional detrapping processes can be present. For example, even in 1D, the de-trapping is more complicated because the wave amplitude and phase velocity are evolving and the wave is not monochromatic. In 2D and 3D, an electron moving across the wave front of a finite-width wave can additionally be detrapped because it leaves the wave. Nevertheless, in all cases, by accelerating across the wave front all trapped electrons will now only extract energy from the wave. 

To study how this nonlinear damping effects SRS,  we carry out one- and two-dimensional (1D and 2D) simulations using the electromagnetic particle-in-cell (PIC) code OSIRIS 4.0 \cite{hemker:thesis}.  The electrons have a temperature $T_e = 3$ keV and slight linear density gradient $n_e = 0.128-0.132 n_{cr}$ ($k\lambda_{D} \approx 0.30$ for backward SRS); ions are fixed to focus solely on SRS interactions.  We simulate an f/8 speckled laser beam of wavelength $\lambda_{0} = 0.351 \mu$m, and in the single-speckle case we emulate a single f/8 speckle with a Gaussian laser beam with focal width $f\lambda_0 = 2.8 \mu$m (intensity full-width half-max) launched from an antenna at the boundary.  The quoted laser intensities are at the focus and range over $6 \times 10^{14} - 5 \times 10^{15}$ W/cm$^{2}$ (where the normalized field of the laser $eE/mc\omega_0 \equiv eA/mc^2=8.5\times10^{-10} \sqrt {I (W/cm^2)} \lambda_{0}(\mu m)= 0.00735 - 0.0212$).  The laser propagates along $\hat{x}$ and is polarized in the 2D plane in $\hat{y}$; $B_{ext}$ is applied in $\hat{z}$, though similar results have been seen by applying $B_{ext}$ in the $\hat{y}$ direction.  The 2D laser profile is only finite-width in $\hat{y}$, so resonant electrons can only be kicked out of the speckle by traveling in the $\hat{y}$ direction.  We used 512 (256) particles per cell in 1D (2D) simulations with cubic interpolation, a grid with 10740 x 1194 cells, and a simulation box of size 120 x 20 $\mu$m$^2$.  The length corresponds to the central portion of an $f/8$ speckle of length $5f^2\lambda_0 = 120 \mu$m.  We simulate approximately 6 ps in time.  The multi-speckle simulations have a width of 42 $\mu$m (approximately 15 speckle widths) with absorbing boundaries for the fields and thermal-bath boundaries for the particles in $\hat{x}$ but periodic boundaries in $\hat{y}$.  For the single-speckle simulations, we use absorbing and thermal boundaries in both $\hat{x}$ and $\hat{y}$ in order to prevent the speckle from interacting with energetic particles and scattered light that would otherwise re-circulate in the transverse direction.  This has the further consequence that exiting particles do not retain their gyro motion when crossing the boundary.  However, single speckle simulations with periodic boundaries show similar features.  Furthermore, the single speckle simulations here are illustrative of the relevant physics and the multi-speckle simulations retain all proper cyclotron motion of the particles.

$B_{ext}$ ranges up to 50 T.  For $B_{ext} = 15-60$ T, the normalized cyclotron frequency $\bar\omega_{c} = 0.001-0.005$.  In a plasma with $T_e = 3$ keV and $T_i = 1$ keV, the Larmor radius for a thermal electron is $r_{e} = 8-2$ microns and for a thermal proton is $r_i = 20-0.5$ millimeters.  The electron (ion) cyclotron period is on the order of a picosecond (nanosecond).  For the single-speckle SRS shown here, the speckle width is on the same order as $r_e$ (several microns) and the time for an e-folding of SRS is on the order of the gyro period (picoseconds).  The ions will execute a gyro period on a time scale much longer than the timescales of interest for the kinetic bursts of SRS.  We have performed several mobile ion simulations and found the conclusions drawn here to be unchanged.

\begin{figure}
\centering
\includegraphics[width=\columnwidth]{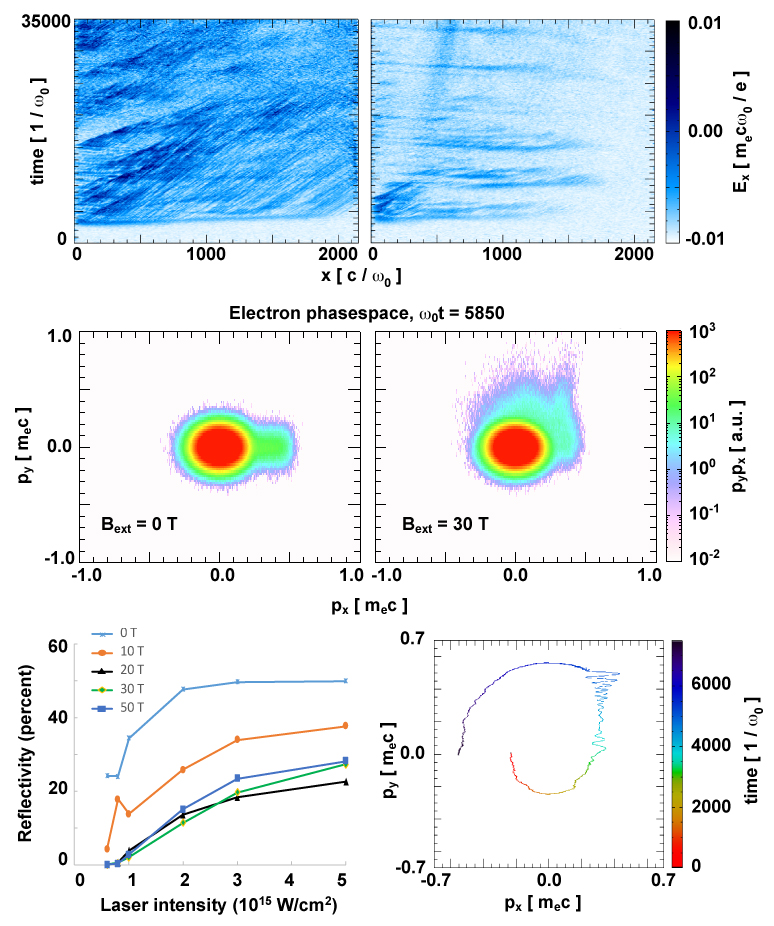}
\caption{\label{fig:1d-thres} (Top) Time vs space plots of EPW activity with $B_{ext}=0$ (left) and 30T (right).  (Middle) Domain-averaged $p_yp_x$ space near the time of initial SRS saturation for $B_{ext}=0$ (left) and 30T (right).  (Bottom) Time-averaged reflectivity vs laser intensity for several magnetic field amplitudes (left) and one particle track plotted in $p_yp_x$ space over $\omega_0t = 0-7500$ (right).}
\end{figure}

The ability of an external magnetic field to decrease SRS activity is evident in 1D simulations ($1\frac{2}{2}$, i.e., one spatial but three velocity components).  Figure \ref{fig:1d-thres}-top shows the spatio-temporal behavior of EPWs for 1D simulations with $T_e = 3$keV, $I_0 = 3\times10^{15}$W/cm\textsuperscript{2}, and $B_{ext} = 0$ and 30T.  For $B_{ext}=0$, strong SRS is seen with the growth and convection of EPWs over most of the simulated domain and time;  SRS EPW amplitudes in this regime can be on the order of $\epsilon \approx 0.1$.  With $B_{ext} = 30$T, on the other hand, the EPW behavior is much more limited in time for each burst of SRS, the EPW peak amplitudes are slightly lower, and the total time-averaged reflectivity level is decreased.  Time-averaged reflectivities across a range of laser intensities and B-field amplitudes are shown in Figure \ref{fig:1d-thres}-bottom.  At the kinetic threshold where SRS just begins to reflect light, $B_{ext}$ decreases the reflectivity to 0.  For larger intensities, the reflectivity can be decreased by at least 50\%.  For constant laser intensity, the reflectivity decreases for increasing $B_{ext}$, though the decrease appears to asymptote and progressively larger $B_{ext}$ are not always able to decrease these 1D reflectivities to 0.

To investigate the mechanism behind this decrease in SRS, we tracked particle orbits.  One representative trapped particle orbit is shown in Figure \ref{fig:1d-thres}-bottom in $p_xp_y$ phasespace, where time is represented by the color-scale.  The particle initially gyrates in the B-field.  It then approaches the EPW phase velocity ($p_x/m_ec \approx 0.3$) and oscillates about this value as it bounces in the wave.  While it is trapped, the electron is accelerated across the EPW wavefront in $p_y$.  Eventually the particle detraps, at which point it continues executing cyclotron motion with a larger energy. The particle gains enough momentum in $\hat y$ that its correct velocity in $v_x$ must be considered relativistically, illustrating that for ICF parameters relativistic corrections need to be included.  In addition, it is accelerated to an energy of approximately 75 keV; if such particles are not confined by the B-field and escape towards the fuel target, they could be a pre-heat threat.  The acceleration of many electrons in such a manner is evidenced in Figure \ref{fig:1d-thres}-middle.  For the case with $B_{ext}=0$, particle acceleration by SRS-generated EPWs is predominantly in the $p_x$ direction and up to a maximum of $p_x/m_ec \approx 0.5$.  For $B_{ext}=30$T on the other hand, trapped electrons with $p_x/m_ec \approx 0.3$ are accelerated in $p_y$ (and accelerated to momenta $>0.5$); once detrapped, these energetic particles gyrate about the field, as evidenced for example by the range of energetic particles with $p_x/m_ec \ll 0.3$ and $p_y/m_ec > 0.3$.  This cross field acceleration mechanism is sufficient to disrupt the nonlinear damping of EPWs during and after SRS saturation and to severely impact the time-averaged behavior of the instability. 

\begin{figure}
\centering
\includegraphics[width=\columnwidth]{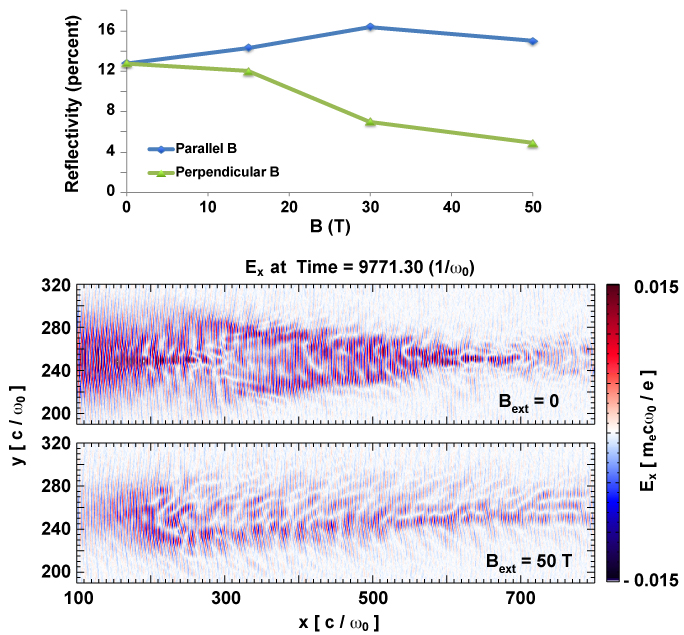}
\caption{\label{fig:one-speckle-e1-xt} (Top) Time-averaged SRS reflectivity as a function of B-field amplitudes. (Bottom) Snapshots in time of 2D EPWs during SRS. }
\end{figure}

We next look at simulations of SRS in single laser speckles with $I_0 = 3 \times 10^{15}$ W/cm$^{2}$.  Figure \ref{fig:one-speckle-e1-xt}-top shows the time-averaged reflectivity as a function of B-field strength for orientations both parallel and perpendicular to the laser $k_0$ ($\hat x$ and $\hat z$ directions, respectively). As the B-field increases in magnitude, the reflectivity decreases significantly for the perpendicular case while increasing slightly for the parallel case.  For these single speckle simulations, the waves have a finite width and the B-field can now not only accelerate trapped particles across the EPW wavefronts but also deflect them out of an unstable region in physical space.  This results in a novel kinetic evolution of finite-width EPWs, as evidenced in Figure \ref{fig:one-speckle-e1-xt}-bottom, where snapshots of nonlinear EPWs for $B_{ext}=0$ and $50$ T are shown.  For $B_{ext} = 0$, the wavefronts are bowed symmetrically about the central axis due to the nonlinear frequency shift on either side of the EPW and the wave is broken up due to the trapped particle modulational instability.  When there is a perpendicular $B_{ext}$, on the other hand, trapped particles traveling in $\hat{x}$ are accelerated in the $\hat{y}$-direction by $B_{ext}$, resulting in nonlinear damping that is different on the top-half of the EPW than on the bottom-half.  The wavefronts are bowed on the bottom half of the spatial domain but, in the top half of the domain, the EPW packet is much lower in amplitude and more disrupted in space.  

Although not shown, for $B_{ext}$ parallel to $k_0$, there is negligible visible difference in EPW behavior outside of what looks like statistical variability, though the EPW activity grows over a longer part of the spatial domain than in the case with $B_{ext} = 0$.  With $B_{ext}$ aligned with the wave vectors of the incident laser and the SRS EPW, the trapped particles gyrate in $\hat y$ and $\hat z$ and, in the absence of relativistic mass corrections, they still execute normal bounce oscillations in the parallel direction.  Since they are more strongly confined to the speckle region, there is less trapped-particle side-loss and the SRS ends up being more 1D-like. This can give more SRS and higher reflectivity, though here the gyroradius of an electron moving at the phase velocity is larger than the speckle width and the increase in SRS is relatively slight.  

\begin{figure}
\centering
\includegraphics[width=\columnwidth]{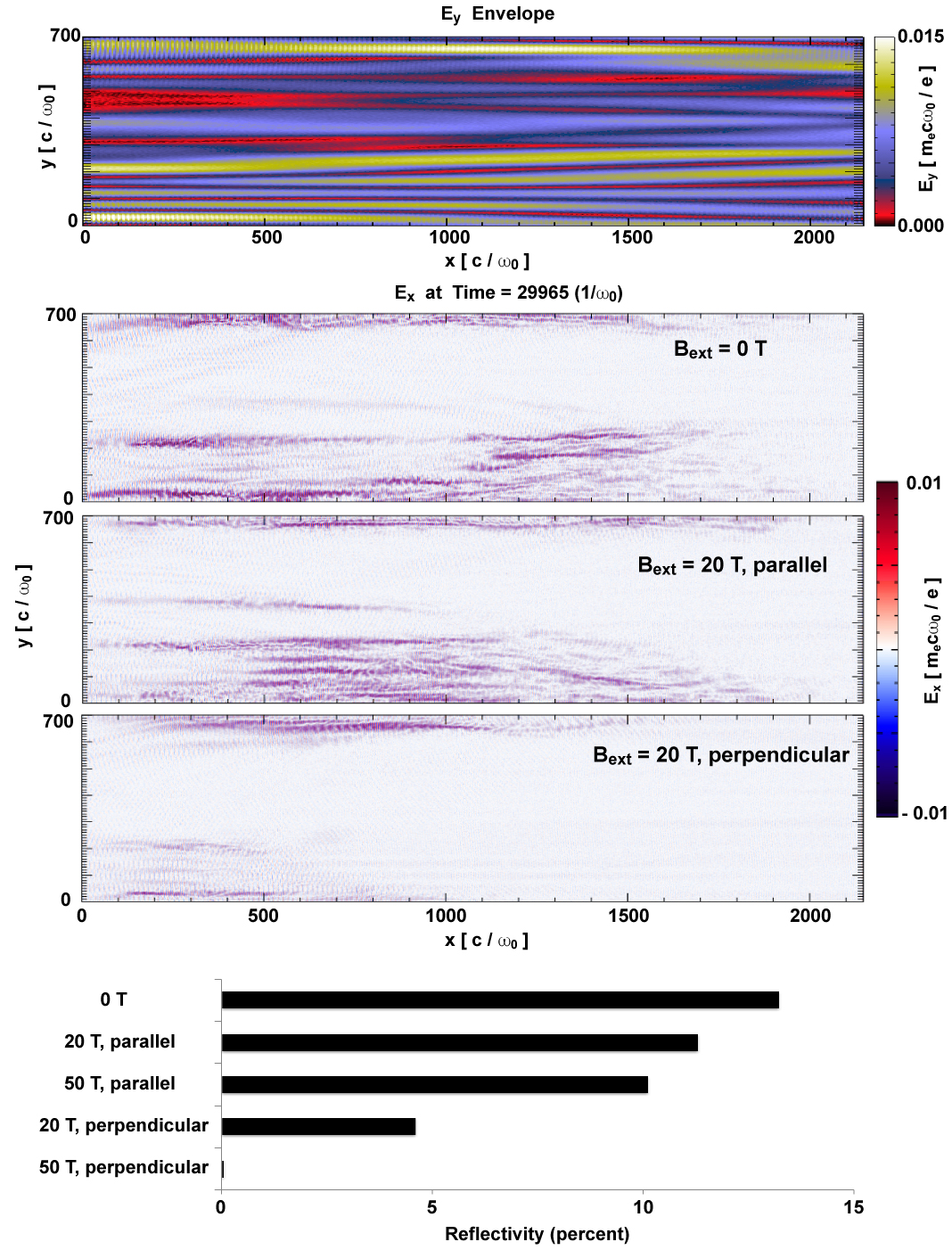}
\caption{\label{fig:multispeckle} (top) Representative snapshot of the incident laser envelope early in time.  (middle) Snapshots in time of EPW activity for $B_{ext} = 0$, 20T parallel, and 20T perpendicular; (bottom) Time-averaged reflectivities for different $B_{ext}$.}
\end{figure}

Finally, we simulated SRS in a multi-speckled laser beam with $I_{ave} = 8 \times 10^{14}$ W/cm$^{2}$. The incident laser profile is shown in Figure \ref{fig:multispeckle}-top. SRS in multi-speckled laser beams can grow as a collective phenomenon due to the spray of waves and particles out of an SRS-unstable region \cite{yin:12,yin:13,winjum:18}.  Consequently, the effect of $B_{ext}$ on multi-speckled SRS depends not just on its influence on single bursts but also on its effect on how waves and particles generated in one burst can travel into other regions that have not yet become unstable.  While we have shown that $B_{ext}$ aligned with $k_{0}$ can act to enhance SRS activity, it also limits the transverse motion of trapped particles, and this in turn limits collective multi-speckle SRS (as was hypothesized by Yin \textit{et al.}~\cite{yin:13}).  Figure \ref{fig:multispeckle}-middle shows it is difficult to distinguish the EPW activity generated by SRS bursts in a case with no B field versus that with a 20 T field aligned with $k_0$. 

For $B_{ext}$ perpendicular to $k_0$, on the other hand, the decrease in plasma wave activity (fourth plot from top) and reflectivity (bottom plot) is much more pronounced.  This appears to be due to several reasons. First, the crossed B-field can prevent an undamped EPW from forming, thereby greatly reducing the number of speckles that are above the laser intensity for kinetically inflated activity.  Second, SRS activity in above-threshold speckles is reduced by EPW damping in the crossed B-field.  Third, the impact of SRS from above-threshold speckles on neighboring speckles is reduced, both because their production of scattered light waves and trapped particles is reduced and because it is more difficult to trigger SRS in neighboring below-threshold speckles since they are ``further'' from threshold.  Finally, the spatial range of trapped particles is confined more closely to existing regions of instability by the cyclotron motion due to $B_{ext}$.

While we have shown that magnetic fields may dramatically affect the evolution of SRS, changing the threshold of SRS in a density gradient may make SRS grow predominantly at higher densities, or lower $k\lambda_D$, and SRS in higher density regions may have higher saturation levels.  Furthermore, B-fields could potentially increase (rather than decrease) SRS by interfering with the nonlinear frequency shift and limiting the effect of detuning which can saturate SRS.  The most realistic angles for B-fields inserted into a hohlraum may be aligned along the hohlraum axis, placing the beams with the highest levels of SRS at approximately 32 degrees relative to the B-field. Bandwidth (ISI \cite{obenschain:89} and SSD \cite{macgowan:96}) and/or STUD pulses \cite{afeyan:13,huller:13,albright:14} combined with magnetic fields may work well, as the EPW may dissipate more strongly during times when the laser is "off" at some spatial location.  Finally, other instabilities may be affected by B fields, such as the two plasmon and high frequency hybrid instability. The kinetic evolution of nonlinear plasma waves in weakly magnetized plasmas is therefore a ripe area for research. 

This work was supported by DOE under Grant No. DE-NA0002953 and the NSF under Grant No. ACI-1339893. Simulations were performed on the UCLA Dawson2 Cluster, NSF's BlueWaters, and ALCF's Mira.

\end{document}